\documentclass[twocolumn,showpacs,aps]{revtex4}

\usepackage{graphicx}  
\usepackage{dcolumn} 
\usepackage{bm} 

\input psfig.sty

\def\bea{\begin{eqnarray}}
\def\eea{\end{eqnarray}}
\def\ben{\begin{equation}}
\def\een{\end{equation}}
\def\benu{\begin{enumerate}}
\def\enu{\end{enumerate}}

\def\n{n}

\def\sss{\scriptscriptstyle\rm}

\def\g{_\gamma}

\def\l{^\lambda}

\def\1var{(\bx_1...\bx\N)}

\def\half{\frac{1}{2}}

\def\br{{\bf r}}

\def\bx{{x}}

\def\x{_{\sss X}}
\def\c{_{\sss C}}

\def\xc{_{\sss XC}}

\def\N{_{\sss N}}

\def\pot{^{\rm pot}}

\def\PBE{^{\rm PBE}}

\def\pot{^{\rm pot}}

\def\ee{_{\rm ee}}

\def\sph_int{ {\int d^3 r}}

\def\JCP{J. Chem. Phys.\ }

\input rotate 
\input epsf

\def\lam{\lambda}

\def\n{n}

\def\wf{\Psi_n^{min ,\lambda }}

\def\V{\hat{V}} 
\def\T{\hat{T}}

\def\winf{U\xc(\infty)} 
\def\winfp{U\xc^\prime(\infty)}
\def\winfc{U\c(\mu=0)} 
\def\winfpc{U\c^{\prime}(\mu=0)}
\def\winfx{U\x(\infty)} 

\def\winfpbe{U\xc^{PBE}(\infty)} 
\def\winfcpbe{U\c^{PBE}(\infty)}

\def\wc{\winfc} 
\def\wx{\winfx} 

\def\pot{v} 
\def\d3r{d^3r\;}

\def\l1{^{\lambda=1}}

\def\cgl{\c^{GL2}}

\def\ucl{U\c(\lambda)}
\def\xcl{\xc(\lambda)}
\def\ucm{U\c(\mu)}

\begin{document} 

\preprint{RUTGERS DFT GROUP: pre-print MTB02}

\title{Accurate Adiabatic Connection Curve Beyond 
the Physical Interaction Strength}

\author{R. J. Magyar}  \affiliation{Department  of  Physics,  Rutgers
University, 136 Frelinghuysen Road, Piscataway, NJ 08854-8019}

\author{W. Terilla} 
\affiliation{Department of Chemistry, Rutgers University, 
610 Taylor Road, Piscataway,
NJ 08854-8019}

\author{K. Burke} 
\affiliation{Department of Chemistry
and Chemical Biology, Rutgers University, 610 Taylor Road, Piscataway,
NJ 08854-8019}


\begin{abstract}
The adiabatic  connection curve of density functional  theory (DFT) is
accurately calculated  beyond the physical  interaction strength for Hooke's atom, 
two interacting electrons in a harmonic  well potential. 
Extrapolation of  the accurate curve  to the 
infinite  coupling limit
agrees well with the  strictly correlated  electron (SCE)
hypothesis but the approach to this limit is more complex.   
The interaction strength interpolation is shown to be a good, but not perfect, fit
to the adiabatic curve.  Arguments about the 
locality of functionals and convexity of the adiabatic 
connection curve are examined in this regime.
\end{abstract}


\draft 
\date{\today} 

\maketitle


\section{Introduction}
\label{s:intro} 

Density functional theory (DFT) is a popular computational
method in solid state physics  and quantum chemistry since 
it is both simple
and reliable \cite{Kb99,HK64,KS65}. 
Because of its wide range of applications and its ability to 
handle large systems, there is considerable interest in DFT 
and improving its accuracy.  
In DFT, the only part of the total energy to approximate is 
the exchange-correlation energy functional, $E\xc[n]$.  
A formal and  general expression for the exchange-correlation energy 
is according to the
adiabatic connection \cite{LP75}, 
\ben 
E\xc [\n] = \int_0^1 d\lambda\;U\xc [n] (\lambda).
\label{adiacon} 
\een
where $U\xc[n](\lambda)$ is the exchange-correlation potential 
energy of a density, $n$,
at coupling constant, $\lambda$. 
Analysis of  the integrand, $U\xc[n](\lambda)$, leads to  many exact
relationships  that  the  exact exchange-correlation energy satisfies 
and approximate functionals should satisfy.  
For example, G{\"o}rling  and Levy obtained a perturbation 
series expression for
the exchange-correlation energy \cite{GL94} by expanding about the weak interaction limit.   
Another fruitful result is  the understanding of
why  hybrid functionals like PBE0  \cite{AB99} and  B3LYP \cite{Bb93}
perform so well \cite{BPEb97,PEB96,BEP97}.   

Because the  exchange-correlation  energy  is  the area  under  
the  adiabatic
connection  curve between  $\lambda=0$  to  $1$,
the most interest in $U\xc (\lambda)$ has been confined to this 
domain.   
However,  there  is  no  fundamental reason  to  restrict
study to this  domain.  In fact, certain exact properties of the 
adiabatic connection curve outside this domain have been used to better 
approximate the curve \cite{PKS1}.  
One example is the consideration of the strong interaction limit, 
$\lambda\rightarrow\infty$.  
A model for this strongly interacting limit is the strictly correlated
electron (SCE) hypothesis \cite{SPL99,S99} 
which states that, because of the strong Coulomb repulsion, 
the individual electrons 
distribute themselves as far apart as possible 
but are constrained to yield a given density.  
Finding one electron uniquely pins the others into position.  
Among other predictions, this SCE model says that
$U\xc$ can also be expanded about the strong interaction strength limit 
($\lambda\rightarrow \infty$). 
Information from this infinite limit combined 
with the G{\"o}rling-Levy expansion about $\lambda=0$ 
leads 
to the suggestion of the interaction 
strength interpolation (ISI) for the entire curve.  
Exchange-correlation energies from the ISI 
are considerably more accurate than those using only  
the first two terms in the perturbation series \cite{SPK0a}. 

Another reason 
to consider large coupling strengths is that approximate exchange-correlation energy 
functionals for this limit might be more accurate\cite{SPK0b}.   
It   has  long  been  known  that standard
approximate  density functionals, such as the local density approximation (LDA) 
or the PBE generalized gradient approximation (GGA),
are better  for exchange-correlation
together  than  they  are  for  exchange alone.  This is due to  a
cancellation  of errors  between  approximations to  the exchange  and
correlation energy  \cite{BEP97,BPL96}.  If this cancellation 
between exchange
and correlation  grows with larger coupling constants, 
approximate density  functionals in this regime will  be   more  accurate.

The present  work is  a detailed study of some of these
suggestions.   We employ a procedure developed  for the range
$\lambda=0$  to $1$  \cite{FTB0} and extend  the simulated  adiabatic
connection curve to larger coupling constants.  
At some point along the adiabatic 
connection curve, the simulating scaling method is expected to break down.
Nevertheless, the curve can be extrapolated from there to the 
infinite coupling limit.  This analysis yields interesting new information 
about the strong interaction limit.

We  work with Hooke's  atom because  
it remains bound no matter now strongly the electrons interact.
Hooke's atom  is the unpolarized two electron system described 
by the Hamiltonian \cite{KHHM93}
,
\bea
\hat{H} = -\half \left( \nabla_1^2 + \nabla_2^2 \right) 
+\frac{k}{2} \left(\br_1^2+\br_2^2\right) 
+ \frac{1}{|\br_1-\br_2|} ,
\label{hooke}
\eea
where $k$ is the harmonic force constant,
$\br_1$ and $\br_2$ are the position operators for each electron, and 
$\nabla_1^2$ and $\nabla_2^2$ are the Laplacian operators for each.
Throughout,  we  use  atomic  units ($e^2=\hbar=m_e=1$)  so  that  all
energies are in Hartrees and all lengths in Bohr radii.
This is not just an exactly solvable
model with  Coulomb interactions but also  an important \emph{physical
system}.  For example, many  authors have  used this
system to model  quantum dots \cite{MHW91,GCL98}.  

Although we  could have  performed  calculations  for  the Hooke's  atom  
at  various harmonic well strengths, we will focus on $k=1/4$.  For this
spring  constant,  the  Hooke's  atom  happens to  admit  an  analytic
solution  \cite{T93}.  
Furthermore, for this $k$ value, the correlation energy is comparable to that of the  Helium atom.

The simulated curves indicate that the SCE predictions for $\winf$ are correct.  
Next, assuming the validity of the SCE hypothesis, we generate a highly accurate simulation of 
the entire curve.
This allows us to
calculate higher 
derivatives of $U\xc(\lam)$ around key points: $\lambda=0$, $1$, and 
$\infty$.  
 This information should be useful for the testing 
and improvement of existing functionals.
We also compare the interaction strength interpolation (ISI) with 
the accurate simulated result.  


\section{Adiabatic Connection Theory}
\label{s:theory}

Three theoretical elements are vital to the content of this paper.  
These are the adiabatic connection curve, the strong coupling limit, 
and the relationship between scale 
factor and coupling constant.  

First, we review the adiabatic connection formalism.  
The integrand of Eq. (\ref{adiacon}) is 
\ben 
U\xc[n] (\lambda) = \langle \wf| \V\ee | \wf \rangle -U[\n] ,
\label{uxc} 
\een
where $U[\n]$ is the Hartree energy, $\V\ee$ is the 
electron-electron Coulomb interaction, and
$\wf$ is the wave-function that 
minimizes  
$\langle \wf| \T+\lambda \V\ee | \wf \rangle$ 
and yields the density $\n (\br)$.
The functional, $U\xc[n] (\lambda)$,
as a function of $\lambda$ makes up the adiabatic connection curve.  
At $\lambda=0$, Eq. \ref{uxc} is just $E\x$, 
the exchange energy evaluated at a given density.  
Later for convenience, we will subtract this contribution and write   
$U\c(\lambda) = U\xc(\lambda)-E\x$.

At small $\lambda$, one may write the  G{\"o}rling-Levy perturbation series \cite{GL92}:
\bea
U\xc[\n] (\lambda)=  E\x[\n] + 2  E\cgl[\n] \lambda +  {\cal O}(\lambda^2)
, 
\;\lambda\rightarrow 0
\label{glps}  
\eea 
where $E\x$ is the exchange energy, and $E\cgl[\n]$ is the first 
order contribution to the correlation energy.
To get the exchange-correlation energy 
from Eq. (\ref{glps}), we need to integrate from $\lambda=0$ to $1$.
Unfortunately, there is no guarantee that the 
higher order terms will be negligible and that the series will converge
\cite{SPK0a}.     

Other exact properties of $U\xc$ might be useful to help 
understand this curve.
An interesting  limit is when $\lambda\rightarrow\infty$.  
This leads us to the second theoretical point, the strong coupling limit.
This limit corresponds to  strongly interacting  electrons which
still yield the physical density.   In this  limit, the integrand
is finite \cite{S99}. 
We can expand $U\xc(\lam)$ about the infinite limit:
\ben
U\xc [\n] (\lambda) = U\xc [\n](\infty) + U\xc' [\n](\infty) /\sqrt{\lambda}
+ {\cal O}\left(1/ \lambda\right),
\;\lambda\rightarrow \infty
\label{winfp}
\een
where 
$U\xc [\n](\infty)$ 
and
$U\xc' [\n](\infty)$ 
are the zeroth and first terms in the expansion.
It has been suggested that the electrons behave in a strictly
correlated manner at this limit \cite{S99}.  
The electrons still produce a given density 
distribution, but finding one electron determines the position 
of all the others. 
Information about this limit can be incorporated into an 
interpolation formula which reproduces both limits 
exactly and can be integrated analytically.   An example is 
the Interaction Strength Interpolation (ISI) \cite{SPK0a}. 

For spherically symmetric two-electron systems in three dimensions, 
the SCE model admits an exact solution for $\winf$ and provides one of two 
contributions to $\winfp$ \cite{S99}.  
One question asked in this paper is how large 
the missing contribution to $\winfp$ is.
We have 
calculated the SCE limit and part of the first correction term
for the Hooke's atom $k=1/4$ 
according to the formulae given by Seidl in Ref. \cite{S99}.  

The final point, the relationship between coupling constant and scale factor, 
is important for the procedure we used to simulate the adiabatic 
connection curve.  A density, $\n(\br)$, is scaled according to 
\bea
\n\g (\br) &=& \gamma^3
\n(\gamma\br),~~~~0 \leq \gamma < \infty.
\label{scaling_law}
\eea
with $\gamma$ being the scale factor.
The exchange-correlation energy at a coupling constant, $\lambda$, and density, $\n(\br)$, 
is simply related to the exchange-correlation energy at a scaled density \cite{B97,LP85}:
\ben 
E\xc^\lambda[\n] = \lambda^2 E\xc[\n_{1/\lambda}] .
\label{coupling_scaling}
\een
The integrand in Eq. (\ref{adiacon}) is $U\xc(\lambda)=d E\xc^\lambda / d\lambda$.
Under  both coupling  constant  and
scaling transformations, we can sometimes show how parts of the 
exact energy transform.  For example,
\ben
E\x^\lambda[\n] = \lambda E\x[\n] 
\mbox{   or   } 
E\x [\n_\gamma] = \gamma E\x[\n] .
\label{ex}
\een
We use this observation later to identify scale factors 
between two scaled densities.


\section{Simulated Scaling Method}
\label{s:method}

In order to generate highly accurate adiabatic connection plots, 
we use the procedure developed  by Frydel, Terilla, and Burke  \cite{FTB0}.  
To find the adiabatic connection curve, we need $E^\lambda\xc[n]$ 
for a set of $\lambda$'s.  For Hooke's atom, 
we know the exact densities and the exact $E\xc$ at different $k$ values.  
Instead of changing $\lambda$, which is difficult, 
we use Eq. (\ref{coupling_scaling}).
A 
small change  in the strength  of external potential yields another density, 
qualitatively  similar  to  the original  density  but on  a
different scale.  If we can solve the system exactly at this different
external potential  strength, 
we have an approximation  to the exchange-correlation energy
with  a scaled  density.  For  densities that  do  not qualitatively
change shape much, this scheme  is highly accurate. 
\begin{figure}[t]
\unitlength1cm
\begin{picture}(12,6.5) 
\put(1.0,0.5) 
{\psfig{figure=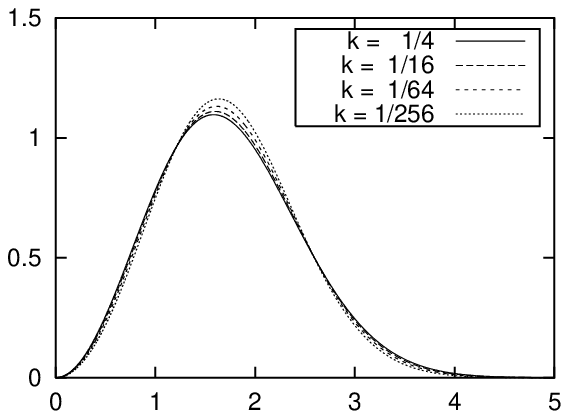,width=8cm,height=7cm}}
\setbox6=\hbox{\large $4\pi r^2 \n'(r)$}
\put(0.5,3.8){\makebox(0,0){\rotl 6}}
\put(4.5,0.0){\large $r$}
\end{picture}
\caption{Simulated scaling of the density.  
We start with Hooke's atom at $k=1/4$.  
Then, we solve at various other coupling constants 
and use the simulated scaling to return us 
as closely as possible to the $k=1/4$ density.}
\label{f:densities} 
\end{figure}
\begin{table}[b]
\begin{center}
\caption{\label{t:lam_and_k} Simulated scaling $k$ and $\lambda$ equivalences 
using the $E\x$ scaling rule, Eq. (\ref{getlam}), to determine $\lambda$.}
\begin{tabular}{|lclc||lclc|} 
\hline
k      & $\lambda$   &&&&&  k      & $\lambda$  \\ \hline
1/4    & 1.000       &&&&&  1/4    & 1.000     \\
1/16   & 1.460       &&&&&  1      & 0.689     \\
1/64   & 2.151       &&&&&  4      & 0.478     \\
1/256  & 3.197       &&&&&  16     & 0.334     \\ \hline
\end{tabular}
\end{center}
\end{table}
To find $U\xcl$, we  differentiate  Eq. (\ref{coupling_scaling}) 
with this highly accurate 
approximation to  the exact $E\xc[\n_{1/\lambda}]$.
Including a  first order
correction  term increases the accuracy of this method:  
\bea 
E\c  [\n\g] &\approx  & E\c  [\n'] +  \int \d3r
\pot\c [\n'](\br) \left(\n\g(\br)-\n'(\br)\right) \nonumber \\ &+& {\cal O}
(\delta \n)^2 ,
\label{hauptrelation}
\eea
where $\pot\c(\br)= \delta E\c[n] / \delta n(\br)$  is 
the correlation contribution to the Kohn-Sham potential. 
The method gives highly accurate energies 
for Hooke's atom ($k=1/4$) and for Helium when 
$\lambda$ varies from $0$ to $1$. 
The error at $\lambda=0$ is $0.3$ mHartrees, and the estimated error 
for $\lambda$ close to one less than 1 mHartree \cite{FTB0}.  

For each simulated scaling, we must assign an appropriate scale factor, 
but which true scaled density does the approximately scaled 
density mimic?  The original paper discusses several possibilities.  
They all require knowing how a chosen component of the energy changes 
with uniform density scaling. 
We use the $E\x$ method:
\ben
\lambda = 1/\gamma = E\x[\n] / E\x[\n'] .
\label{getlam}
\een
Since we use $E\x$ to assign
$\lambda$, the $\wx$ contribution to $\winf$ 
necessarily scales properly for all values of $\lambda$, and so
we show only $\wc$.

In this paper, we examine the adiabatic connection curve 
at large interaction strengths.  
This method only works for $\lambda>1$ for systems that remain bound as the external potential is weakened.  
Even with this restriction,
the method 
must
ultimately fail as $\lambda\to \infty$.  
Specifically for Hooke's atom, Cioslowski showed that at a certain
critical strength for the external potential,
$k_c=0.0016$ ($\lambda_c=4.138$), the density
changes shape qualitatively \cite{CP0}.  
Beyond this value, the simulated scaling might no longer 
be a good approximation to exact scaling.
On the other hand, the method fails for He 
almost immediately as the two electron ion unbinds 
at nuclear charge, $Z=0.9$.

\begin{figure} [ht]
\unitlength1cm
\begin{picture}(12,7.5) 
\put(1.0,0.5)   
{\psfig{figure=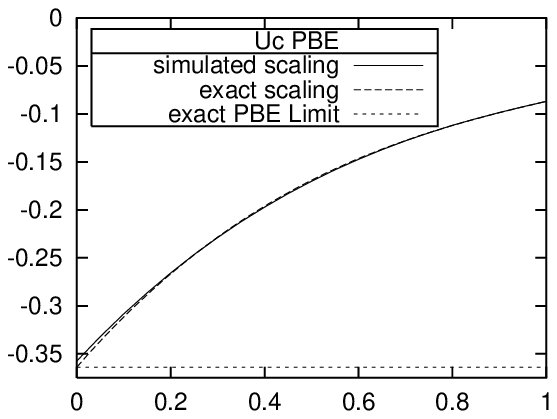,width=8cm,height=7cm}}
\setbox6=\hbox{\large   $U\c(\mu)$}  \put(0.5,3.8){\makebox(0,0){\rotl
6}} \put(4.5,0.0){\large $\mu$}
\end{picture}
\caption{PBE adiabatic connection curve for 
Hooke's  atom
($k=1/4$): $U\c(\mu)$.  
The solid line is generated using simulated scaling of the density, 
and the dashed curve by exactly scaling the known functional.
The exact PBE $\winfc$ limit is shown (short dashes).
}
\label{f:pbe} 
\end{figure}
To test this procedure and to
develop a rule for its reliability, we apply the procedure 
in a case where we already know
the correct answer, namely with an approximate functional.
A generalized
gradient approximation (GGA) mimics the
complexity of the true functional better than, say, the local density approximation.   
Because of its first principle derivation and reliability, 
we use PBE here \cite{PBE96}.  Since we have the analytic form for 
the PBE functional, we can scale the input density  
to generate the entire adiabatic curve, Fig. \ref{f:pbe}. 
The curve is shown 
as a function of  $\mu=1/\sqrt{\lambda}$ 
so that the region $\lambda\in 1,\infty$ can appear on a finite sized plot.

PBE results for certain key $\lambda$ values are listed in Table \ref{t:exactresults}.
An  explicit  formula \cite{SPK0b} for the  PBE  functional as  $\mu
\rightarrow  0$  is        
\bea         
\winfpbe [\n]=
\int d^3r \; \n(r) \epsilon\x(\n) 
\left(F\x\PBE(s)+\frac{0.964}{1+y+y^2} \right) \nonumber \\
\nonumber\\
\label{winfPBE}
\eea 
where $y=0.2263\; s^2$, $s$ is the reduced gradient, 
$\epsilon\x(\n)$ is the exchange energy per particle of the uniform gas,
and $F\x\PBE(s)$ is an exchange enhancement factor \cite{PBE96}.  

We need a criterion for how far along the adiabatic connection 
we can trust the simulated density scaling to mimic the exactly scaled density.  
Our criterion is to terminate 
the simulations at
$\mu=\mu_c=1/\sqrt{\lambda_c}$ where the density 
qualitatively changes shape \cite{CP0}. 
Even at this point, 
the first order correction in Eq. (\ref{hauptrelation}) still improves upon the zeroth order simulation.
This  is a
highly  conservative estimate;  it  is likely that the curves 
are accurate to 
smaller $\mu$'s.  

To get a prediction for $\winfc$, we must extrapolate 
the simulation to $\mu=0$.  
This is done by fitting the simulated data to an n$^{th}$ order polynomial 
and extrapolating this polynomial to $\mu=0$.
The third order polynomial connecting four sample points 
best reproduces the known $\winfcpbe$.  
In Fig. \ref{f:pbe}, we show the exactly scaled PBE functional 
and the polynomial interpolation.
We see that the simulated curve is almost on top of the exact curve.  
However, they do differ slightly in the $\winfc$ values.  For the 
simulated curve, 
$\winfc=-0.357$, and the scaled result is 
$-0.363$ from Eq. (\ref{winfPBE}), a 6 mHartree error. 


\section{Extrapolating to the Infinite Interaction Strength Limit}
\label{s:extrap}
\begin{figure} [ht]
\unitlength1cm
\begin{picture}(12,7.5) 
\put(1.0,0.5)   
{\psfig{figure=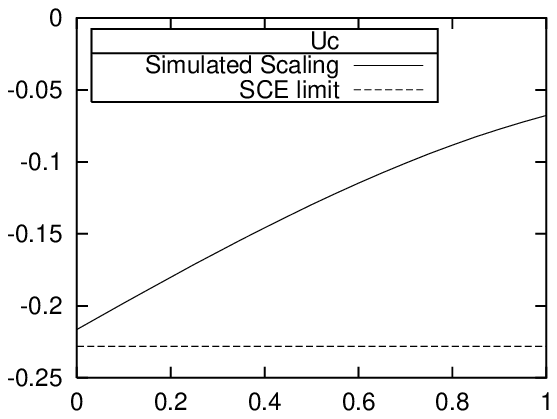,width=8cm,height=7cm}}
\setbox6=\hbox{\large   $U\c(\mu)$}  \put(0.5,3.8){\makebox(0,0){\rotl
6}} \put(4.5,0.0){\large $\mu$}
\end{picture}
\caption{The adiabatic connection  curve for  Hooke's  atom ($k
=1/4$): $U\c(\mu)$.  The solid line is the simulated curve.  
The SCE limit is shown as a dashed line.}
\label{f:winf} 
\end{figure}

The simulated  
adiabatic connection curve for Hooke's atom $k=1/4$ 
in Fig. (\ref{f:winf}) approaches the SCE $\winfc$ limit.  
As in section \ref{s:method} for the PBE functional, 
we reproduce the entire curve by fitting the simulated points to a third order 
polynomial.
Since the simulated scaling method is only reliable between $\mu_c=1/2$ and $1$, 
we must extrapolate the curve over the domain $\mu=0$ to $1/2$ 
by a polynomial.  
The extrapolated 
prediction for $\winfc$, $-0.206$,  
is $22$ mHartrees  from the  strictly 
correlated  electron
prediction, $-0.228$.  
We do not expect as good agreement because the true $E\c$ functional is 
more complicated than a GGA, and we regard
the result as consistent
with the SCE hypothesis.  


\section{Simulating the Entire Adiabatic Connection Curve}
\label{s:simul}
\begin{figure} [h]
\unitlength1cm
\begin{picture}(12,7.5) 
\put(1.0,0.5)   
{\psfig{figure=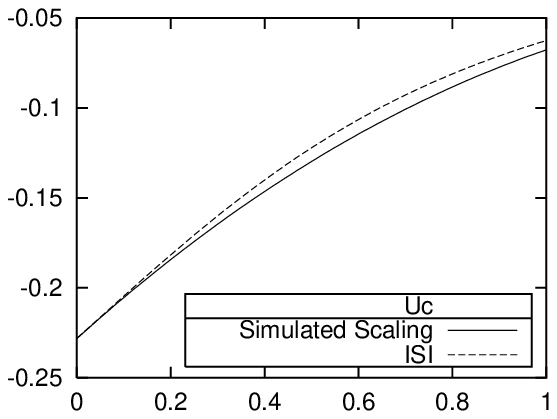,width=8cm,height=7cm}}
\setbox6=\hbox{\large                       $U\c(\mu)$}
\put(0.5,3.8){\makebox(0,0){\rotl  6}}  \put(4.5,0.0){\large $\mu$}
\end{picture}
\caption{Simulated  adiabatic   connection  curve  for   Hooke's  atom
(k=1/4): $U\c(\mu)$.  
The solid line is the simulated curve with the SCE $U\c(\mu=0)$. 
The dashed curve is the ISI using exact inputs.}
\label{f:adiasim} 
\end{figure}

In section \ref{s:extrap}, we used an extrapolation scheme 
to complete the adiabatic curve.  
Here, we combine the simulated part with the SCE electron 
limit to produce a highly accurate adiabatic connection curve for
all coupling strengths.   From this curve, we calculate the first 
terms in Taylor expansions about both $\lambda= 0$ and $1$, and $\mu=0$ and $1$.  
Using these new results, we assess the accuracy of the Interaction 
Strength Interpolation (ISI) with accurate inputs.

\begin{table}[ht]
\begin{center}
\caption{\label{t:derivsm} Higher derivatives of $\ucm$ with respect to $\mu$  
for Hooke's atom ($k=1/4$).}
\begin{tabular}{|lccc|}
\hline
$\mu \;\;\;$   & $\ucm$  & $U\c'(\mu)$ & $U\c''(\mu)$   \\ \hline
0              &  -0.228 & 0.235       &  -0.156        \\ 
1              &  -0.068 & 0.088       &   0.221        \\ 
\hline
\end{tabular}
\end{center}
\end{table}
\begin{table}[ht]
\begin{center}
\caption{\label{t:derivsl} Higher derivatives of $\ucl$ with 
respect to $\lambda$
for Hooke's atom ($k=1/4$).}
\begin{tabular}{|lccccc|}
\hline
$\lambda\;\;\;$ & $\ucl$  & $U\c'(\lambda)$ & $U\c''(\lambda)$ & $U\c^{(3)}(\lambda)$ & $U\c^{(4)}(\lambda)$ \\ \hline
0              &  0.0000  &  -0.101        &    0.095    & -0.107 & 0.124 \\ 
1              & -0.0677  &  -0.044	   &    0.032    & -0.032 & 0.039 \\ 
\hline
\end{tabular}
\end{center}
\end{table}

The $\mu<1$ 
simulated adiabatic connection curve is shown in Fig. \ref{f:adiasim}.  
The curve was generated by fitting the simulated data points from $\mu=0.5$ to $1$ 
and including the SCE $\winfc$ in the point set.  
We used a third order 
polynomial,
the order that best reproduced the 
adiabatic curve for the PBE functional
in section \ref{s:method}.
This curve should be
an excellent approximation to 
the exact curve.  
From the plot, we see that 
the derivative $dU\c(\mu)/d\mu$ is positive everywhere along the adiabatic curve.  
This implies that $dU\c(\lambda)/d\lambda$ is negative, 
and the adiabatic curve is convex.
All calculated $U\c(\lambda)$ curves for $0\leq \lambda\leq 1$ have $dU\c(\lambda)/d\lambda<0$, 
but the inequality has never been generally proven.  
Our result extends this observation to $\lambda\geq 1$ for this system. 

\begin{table}
\begin{center}
\caption{\label{t:exactresults} 
Accurate results for Hooke's atom with $k=1/4$ 
evaluated on the exact densities.}
\begin{tabular}{|lcccccc|} 
\hline
 & $E\x$ & $2 E\cgl$ & $E\c$ & $U\c(\mu=1)$ & $\winfc$ & $\winfpc$ \\ \hline
PBE   & -0.493 & -0.168 & -0.051 & -0.087 & -0.363 &  0.561 \\ \hline
Exact & -0.515 & -0.101 & -0.039 & -0.068 & -0.228 &  0.235 \\ \hline
\end{tabular}
\end{center}
\end{table}
\begin{table}
\begin{center}
\caption{\label{t:hooke12}
Interaction Strength Interpolation Results for Hooke's atom with $k=1/4$. 
\emph{Accurate} and \emph{model} refer to the value of $\winfpc$.
The accurate value is from our simulation and the model is from Seidel's model \cite{S99}.}
\begin{tabular}{|lccccc|} 
\hline
Method      & $\winfpc$ & $U\c(\lambda=1)$  & Error   & $E\c$      & Error   \\ \hline
ISI (accurate)       & 0.235   & -0.063     &  8 \%   & -0.036       &  6 \% \\
ISI (model)          & 0.281   & -0.060     & 11 \%   & -0.035       &  9 \% \\ \hline  
\end{tabular}
\end{center}
\end{table}

Derivatives of  $U\c(\mu)$ are obtained from the coefficients in the polynomial extrapolation.
Two higher derivatives of $U\c(\mu)$ with respect to $\mu$ are shown in table \ref{t:derivsm}.  
Seidl's model for $\winfpc=0.281$ \cite{S99} does not agree with the accurate $\winfpc$.  
This indicates that 
the missing contributions to the SCE $\winfpc$ mentioned 
by Seidl are, at least for this system, not negligible.  

Several higher 
derivatives of $U\c(\lambda)$ with respect to $\lambda$ are listed in table \ref{t:derivsl}.  
Here, we need not restrict ourselves 
to a third order polynomial interpolation because we have 
a dense sampling of data points over the range $\lam=0$ to $4$.  
The higher derivatives reported in terms 
of $\lambda$ are expected to be highly accurate.  

The interaction strength interpolation (ISI), as originally formulated \cite{SPK0a}, 
is an interpolation scheme for the entire adiabatic connection curve. 
It used exact values at $\lambda=0$ and carefully chosen GGA values at $\mu=0$.
We now ask how well the ISI with accurate inputs compares 
to the simulated curve.  
The answer tells us how good the choice of curve in the ISI is.
For the inputs to the ISI, we use the  exact $E\x$ and $E\cgl$ which are 
derivable from the simulated curves in Ref. \cite{FTB0} and are given in Table 
\ref{t:exactresults}.  For $\winf$,
we use the SCE prediction which judging 
from the results in section \ref{s:extrap},
we believe to be exact.  
For $\winfpc$, we input two different values.
the accurate simulated value and Seidel's prediction.  
The results are shown in table  \ref{t:hooke12}.  
The ISI interpolation does not 
perform exceptionally well with accurate inputs as already noticed in \cite{PKS1}.  
For example, the magnitude of $U\xc(1)$ is underestimated by 5 mHartrees.  
This is perhaps a result of 
the way the $\winfpc$ limit is included in the interpolation equation.  
For this system, incorporating the accurate value for $\winfpc$ in the ISI 
does \emph{not} greatly improve its accuracy.

\begin{figure} [h]
\unitlength1cm
\begin{picture}(12,7.5) 
\put(1.0,0.5)   
{\psfig{figure=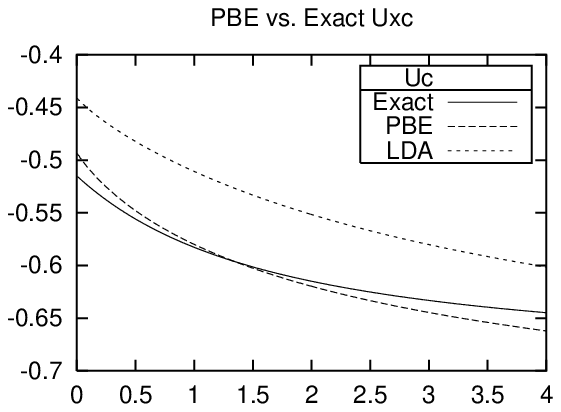,width=8cm,height=7cm}}
\setbox6=\hbox{\large                       $U\xc(\lambda)$}
\put(0.5,3.8){\makebox(0,0){\rotl  6}}  \put(4.5,0.0){\large $\lambda$}
\end{picture}
\caption{Adiabatic   connection  curve  for Hooke's Atom using various functionals: 
The exact curve is the solid line,
the PBE is the long dashed line,  
and  the local density approximation (LDA) is the short dashed line.}
\label{f:pbe_exact} 
\end{figure}

In Fig. \ref{f:pbe_exact}, we see how the PBE and LDA adiabatic 
connection curves compare to the accurate curve.  The PBE curve clearly crosses the accurate curve.  
Since $\lim_{\mu\rightarrow 0}U\xc^{LDA}(\mu) = 1.964 E\x^{LDA} = -0.866 < \winf$, the LDA curve 
must cross the accurate one at some larger interaction strength.
Since both curves cross the exact curve at some $\lambda>1$, the cancellation of errors between exchange 
and correlation in $E\xc^\lambda$ will grow smaller beyond some critical interaction strength and  
become an addition of errors.  
It has been argued that because the exchange correlation on-top hole grows more local
as the interaction strength increases \cite{BEP97,BPE98}, local functionals 
for $E\xc^\lambda$ would work better as $\lambda$ increases.  This is certainly true for our system 
in the range, $0\leq\lambda\leq 1$; however,
the adiabatic plots indicate that as $\lambda$ grows, 
the energy depends on the density in an increasingly nonlocal way.  
The accuracy of the on-top hole is less relevant to the total energies 
in the strongly interacting region of the adiabatic connection curve.


\section{conclusion}
\label{s:conc}

In this work, we have extended the
method of Ref. \cite{FTB0} to simulate the adiabatic 
connection curve to interaction strengths greater 
than the physical value for a simple model system. 
In doing so, we kept in mind that
the method must fail at some $\mu_c$ 
as $\mu \to 0$ ($\mu = 1 /\sqrt{\lambda}$) and performed 
an extrapolation 
to the strong interaction limit.   
This simulated curve agreed with the SCE hypothesis.  To generate a highly accurate 
curve for $\mu=0$ to $1$, we included the SCE $U\c(\mu=0)$ in 
the set of points and interpolated. 
Using this accurate adiabatic curve, we found
higher derivatives at key 
coupling constants: $\lambda=0$,$1$, and $\infty$.  
Finally, we compared some popular approximate functionals to the accurate curve.
These 
results will be useful in the formal analysis of the adiabatic 
connection curve, the testing of approximate functionals, 
and the construction of new functionals in DFT.


\section{Acknowledgments}

We would like to thank John Perdew for discussions 
and Takeyce Whittingham for computationally checking $\winfpbe$.
This work supported by the National Science Foundation 
under grant number CHE-9875091.  



\end{document}